# Second sound waves in diamond

V.D. Khodusov, A.S. Naumovets

Karazin Kharkov National University, Svobody Sq. 4, Kharkov, 61077, Ukraine.

Decrmber 17, 2010

email: vkhodusov@ukr.net

## Abstract

The possibility of propagation of second sound waves in diamond single crystals depending on their dimensions, concentrations of isotopes and temperature is studied. At this correct account of phonon scattering on boundaries is important. The calculation of phonon collision frequencies is carried out in the reduced isotropic crystal model using second and third modules of elasticity and in Callaway model on the basis of experimental data on diamond thermal conductivity. Both models give us the consistent values of parameters under which the propagation of SSW is possible. It is discovered that concentrations of isotopes $^{13}C < 10^{-5}$, temperatures $T < 90K$. Such a good agreement provides the reliability of received results and shows the efficiency of reduced isotropic crystal model in the description of diamond properties in low temperature range.

## 1. Introduction.

In perfect single crystals the propagation of weakly damping second sound waves (SSW), similar to He II ones, is possible [1, 2]. Their propagation in solid bodies is possible when normal processes of phonon interaction with quasi-momentum conservation law are considerably faster than the resistive processes leaking without it (umklapp processes, scattering on impurities, isotopes, boundaries, etc.), i.e. when establishes gas-dynamic regime. In work [1] is shown that such waves might exist not only in phonon gas in solids, but also in arbitrary gas of Boze quasi-particles in gas-dynamic regime.

In work [3] for the first time was found the temperature range of SSW existence in solids, so-called "window", that doesn't appear in liquid helium. Experimentally SSW were observed only in certain crystals: solid helium [4], NaF [5], Bi [6] and sapphire [7]. The major impossibility of SSW monitoring in other crystals is banded with their strong damping on the impurities and isotopes. Thus in perfect pure single crystals the major contribution to SSW damping makes isotopic phonon scattering. Modern technologies give an opportunity to grow not only pure, but also isotopic pure perfect single crystals. The problem of finding such concentrations of impurities and isotopes that do not influence on SSW propagation springs up. This problem can be solved in the reduced isotropic crystal model with respect to the modules of elasticity[1] or Callaway model that describes experimental data on thermal conductivity [8]. In work [1] was built the consecutive theory of finding the kinetic coefficients in Boze quasi-particles gas, that define coefficient of SSW damping. For calculation of kinetic coefficients was offered the reduced isotropic crystal model (RICM) with respect to second and third modules of elasticity.

The interest to possibility of SSW observation in diamond is caused by diamond's highest Debaye temperature, so the "window" of SSW existence may be in considerably higher temperature range than for other crystals in which SSW were observed. In addition, there is experimentally revealed strong dependence of thermal-conductivity coefficient on concentration of isotope $^{13}C$ [8]. In natural diamond concentration of isotope $^{13}C$ is $1.1 \cdot 10^{-2}$, that is significantly higher than concentration of nitrogen impurities, that is on the order of $10^{-5}$. This fact can explain observed strong dependence of diamond thermal conductivity on concentration of isotope $^{13}C$. In synthetic diamonds concentration of nitrogen impurities may reach $10^{-7}$. Further we assume that concentration of nitrogen impurities is considerably lower than concentration of isotopes $^{13}C$ and so we don't take account of SSW damping on these impurities.

Using the experimental values of elasticity modules for diamond from works [9, 10] in present work in RICM were found: dimensions of the sample, temperature window of SSW existence and concentrations of isotopes $^{13}C$ below which the contribution of phonon scattering on isotopes is not essential.

The temperature range of SSW existence lies in the neighborhood of thermal conductivity coefficient maximum in low temperature range [5]. For several crystals the value and location of that maximum considerably depends on isotopic composition. Particularly that reveals in diamond crystals [11, 8]. Such dependence in Callaway model can be explained by a special role of normal processes [11, 12, 13] that are flowing considerably faster than resistive processes, in this case realizes gas-dynamic regime. Owing to it in other works for calculation of the coefficient of SSW damping were used the of phenomenological frequencies of phonon collisions which are employed in Callaway model for the description of experiments on thermal conductivity [14]. In present work the calculations in Callaway model are made for the comparison with results received in RICM. In work [15] was attempted the studying of possibility of SSW existence in diamond using Callaway model, however found in this work coefficient of SSW damping was of the order of frequency, that prejudice the possibility of their experimental detection. In present work is shown that large coefficient of SSW damping in work [15] is conditioned by wrong account of frequency of phonon scattering on boundaries. Right account of these processes shifts the "window" of SSW existence in lower temperature range. Also the usage of two models and good accordance of both results provides high reliability of received results and the efficiency of RICM in the description of physical properties of diamond in low temperature range.

Resonant responses of Mandel'shtam-Brillouin scattering were recently registered by four-photon spectroscopy in the optical glass K8 [16] at a frequency that was attributed to induced generation of second sound wave. A similar phenomenon can be observed in diamond crystals in the region of the second sound waves, which is defined in this paper.

## 2. Gas dynamics equations of Boze quasi - particles.

In the beginning we will obtain the equations of gas-dynamics which describe an arbitrary Boze quasi-particles then we will apply received equations to the description of phonons in solids.

In the kinetic theory the state of quasi-particles gas is characterized by quasi-particles distribution function $N = N^{(j)}(p, r, t)$, which satisfies kinetic equation that has a form of Boltzmann equation:

$$\left(\frac{\partial}{\partial t} + \mathbf{g}\frac{\partial}{\partial \mathbf{r}} - \frac{\partial \varepsilon}{\partial \mathbf{r}}\frac{\partial}{\partial \mathbf{p}}\right) N = \left(\dot{N}\right)_{coll} \tag{1}$$

where $\mathbf{g} \equiv \mathbf{g}^{(j)} = \frac{\partial \varepsilon^{(j)}}{\partial \mathbf{p}}$ – is the group velocity of quasi-particles; $\varepsilon \equiv \varepsilon^{(j)}(\mathbf{p},\mathbf{r},t)$ – is Hamiltonian of quasi-particle, which is equal to its local energy; $(N)_{coll}$ – is the collision integral of quasi-particles, the rate of distribution function change due to different processes of quasi-particles interactions.

The condition of N-processes prevalence can be represented as next inequality:

$$\tau_N \ll \tau_R \tag{2}$$

where $\tau_N$ - the normal process relaxation time, $\tau_R$ - resistive relaxation time.

If in some instant of time the system of quasi-particles goes out of its equilibrium state, then during $T_N$ establishes quasi-local balance. It is characterized by distribution function $N_0^{(j)}$, which turns to zero the collision integral due to N-processes [1]:

$$N_0^{(j)} = \left(\exp\frac{\varepsilon^{(j)} - (\mathbf{pu})}{T_0(1+\theta)} - 1\right)^{-1} \tag{3}$$

where $u$ is drift velocity in gas of quasi-particles, $\Theta = (T - T_0)/T_0$ is relative temperature.

In sub-equilibrium statistical state, the solution of Eq. (1) is sought in the form:

$$N^{(j)} = N_0^{(j)} + \delta N^{(j)}, \quad \left(\left|\delta N^{(j)}\right| \ll N_0^{(j)}\right) \tag{4}$$

where $N_0^{(j)}$ is locally equilibrium distribution function, that depends on gas-dynamic quantities and $\delta N^{(j)}$ depends on theirs gradients. If one work in the thermodynamic potential as:

$$F_0 = -T \sum_{\mathbf{k},j} \ln\left(1 + N_0^{(j)}\right) \tag{5}$$

which satisfies thermodynamic relation as the function of $T$ and $u$:

$$dF_0 = -S_0 dT - \mathbf{P} d\mathbf{u} \tag{6}$$

we can define the momentum density $P$, heat capacity $C$ and entropy $S_0$ densities, and components of quasi-particles density tensor $\tilde{\rho}_{ij}$:

$$\mathbf{P} = -\left(\frac{\partial F_0}{\partial \mathbf{u}}\right)_{T,\tilde{A}}, \quad S_0 = -\left(\frac{\partial F_0}{\partial T}\right)_{\mathbf{u},\tilde{A}}$$

$$C = T\left(\frac{\partial S_0}{\partial T}\right)_{\mathbf{u},\tilde{A}}, \quad \overline{\tilde{\rho}}_{il} = \frac{\partial \overline{P}_i}{\partial u_l} = -\overline{\left(\frac{\partial^2 F_0}{\partial u_i \cdot \partial u_l}\right)}_{T,\tilde{A}} \tag{7}$$

Applying the standard procedure [1] from the kinetic equation (1) we derive gas-dynamic system of equations, which describes an isotropic gas of quasi-particles in the linear approximation to drift velocity $u$ in the model of reduced isotropic crystal:

$$\begin{cases} \dot{\mathbf{P}} + ST\nabla\theta = -r\mathbf{u} + \tilde{\eta}\Delta\mathbf{u} + \left(\tilde{\zeta} + \frac{\tilde{\eta}}{3}\right) \cdot \nabla div\mathbf{u} \\ C\dot{\theta} + S div\mathbf{u} = \tilde{\kappa}\Delta\theta \end{cases} \tag{8}$$

where $\mathbf{P} = \tilde{\rho}\mathbf{u}$, $\tilde{\rho}$ is quasi-particle density, $r$ is coefficient of contact friction, that is conditioned by the quasi-particle processes flowing without quasi-momentum conservation (umklapp processes, quasi-particle scattering on impurities, isotopes and boundaries of the sample), $\tilde{\eta}$, $\tilde{\zeta}$ are first and second viscosities, $\tilde{\kappa}$ is thermal conductivity coefficient.

## 3. Reduced isotropic crystal model (RICM).

In the low temperature range $T \ll \Theta_D$ ($\Theta_D$ – Debaye temperature) long-wave acoustic phonons play fundamental role in thermodynamic and kinetic characteristics of crystals. One can find the dispersion of acoustic phonons and amplitude of their interaction using the macroscopic equations in the elasticity theory of crystals with deformation energy including third order anharmonicity. The crystal energy density in this case is:

$$E = \frac{1}{2}\lambda_{iklm}u_{ik}u_{lm} + \frac{1}{6}\lambda_{ii'kk'll'}u_{ii'}u_{kk'}u_{ll'} \tag{9}$$

where $u_{ik} = \frac{1}{2}\left(\frac{\partial \xi_i}{\partial x_k} + \frac{\partial \xi_k}{\partial x_i} + \frac{\partial \xi_l}{\partial x_i}\frac{\partial \xi_l}{\partial x_k}\right)$ – deformation tensor, $\vec{\xi}$ – displacement vector, $\lambda_{iklm}$, $\lambda_{ii'kk'll'}$ – fourth and sixth rank elastic modules tensors.

In the elastic waves theory in crystals with undefined symmetry underlies Kristoffel's equation:

$$\left(k^2\lambda_{ijlm}\hat{n}_j\hat{n}_l - \omega^2\delta_{im}\right) = 0 \tag{10}$$

where $k$ – propagation vector, $\hat{n} = k/k$, $\omega$ – frequency. Decidability condition of Eq.(10) gives the dispersion equation for the frequencies of acoustic phonons:

$$\det\left(k^2\lambda_{ijlm}\hat{n}_j\hat{n}_l - \omega^2\delta_{im}\right) = 0 \tag{11}$$

In the case of arbitrary vector $\hat{n}$ directions the solution of this equation represents a complicated task. It is possible to use the reduced isotropic crystal model (which elastic properties will be close to the elastic properties of considering crystal) to simplify the calculations of the dispersion laws. Such model is called "the reduced isotropic crystal model" relatively to second order elastic modules [1].

In general case in RICM second order elasticity modules tensors have the following view:

$$\lambda^{(0)}_{ijlm} = \lambda_1\delta_{ij}\delta_{lm} + \lambda_2(\delta_{il}\delta_{jm} + \delta_{im}\delta_{jl}) \tag{12}$$

The tensors of elastic modules of reduced isotropic crystal are found from the condition of minimality of expression $\left|\lambda_{ijlm} - \lambda^{(0)}_{ijlm}\right|^2$ relatively to $\lambda_1$ and $\lambda_2$. For cubic syngony crystals $\lambda_1$ and $\lambda_2$ are equal respectively:

$$\lambda_1 = \frac{1}{5}(c_{11} + 4c_{12} - 2c_{44}), \quad \lambda_2 = \frac{1}{5}(c_{11} + 3c_{44} - c_{12}) \tag{13}$$

where $C_{nm}$ – the second order elastic modules in generally accepted matrix form.

For the reduced isotropic crystal relatively to second order elastic modules the phase velocities of longitudinal (l) and transverse (t) acoustic phonons, theirs relation and average velocity of sound are equal to:

$$V_l = \left(\frac{\lambda_1 + 2\lambda_2}{\rho}\right)^{1/2}, \quad V_t = \left(\frac{\lambda_2}{\rho}\right)^{1/2},$$

$$\beta = \frac{V_l}{V_t} = \left(\frac{\lambda_2}{\lambda_1 + 2\lambda_2}\right)^{1/2}, \quad V_S = V_t\left(\frac{3}{2+\beta^2}\right)^{1/3}. \tag{14}$$

where $\rho$ – crystal density.

Third order elastic tensor in the reduced isotropic crystal is defined by three independent constants *A, B* and *G* called Landau third order elastic modulus. They can be written as:

$$\lambda^{(0)}_{ii'kk'll'} = A\delta_{ii'}\delta_{kk'}\delta_{ll'} +$$
$$+ B[\delta_{ii'}(\delta_{kl'}\delta_{k'l} + \delta_{kl}\delta_{k'l'}) + \delta_{kk'}(\delta_{il'}\delta_{i'l} + \delta_{il}\delta_{i'l'})$$
$$+ \delta_{ll'}(\delta_{ik'}\delta_{i'k} + \delta_{ik}\delta_{i'k'})] + C[\delta_{il}(\delta_{ki}\delta_{k'l'} + \delta_{kl'}\delta_{k'i'}) +$$
$$+ \delta_{il'}(\delta_{ki}\delta_{k'l} + \delta_{kl}\delta_{k'i'}) + \delta_{i'l}(\delta_{ik}\delta_{kl'} + \delta_{ik}\delta_{k'l'}) +$$
$$+ \delta_{i'l'}(\delta_{ki}\delta_{lk'} + \delta_{ik}\delta_{ik'})] \qquad (15)$$

The sixth rank elastic modules tensors of the reduced isotropic crystal, which elastic characteristics are close to the considering crystal are defined from the condition of minimality of expression $\left|\lambda_{ii'kk'll'} - \lambda^{(0)}_{ii'kk'll'}\right|^2$ relatively to *A*, B and *G*. As the results for the cubic syngony crystals the values of these coefficient are equal to:

$$A = \frac{1}{35}(c_{111} + 18c_{112} + 16c_{123} - 30c_{144} - 12c_{155} + 16c_{456})$$
$$B = \frac{1}{35}(c_{111} + 4c_{112} - 5c_{123} + 19c_{144} + 2c_{155} - 12c_{456}) \qquad (16)$$
$$C = \frac{1}{35}(c_{111} - 3c_{112} + 2c_{123} - 9c_{144} + 9c_{155} + 9c_{456})$$

where $c_{ikl}$ -are the third order elastic modulus in generally accepted matrix form.

We will work in the dimensionless coefficients of third order elastic modules in the following way:

$$\overline{a} = \frac{A}{\rho V_l^2}, \overline{b} = \frac{B}{\rho V_l^2}, g = \frac{G}{\rho V_l^2} \qquad (17)$$

In work [1] for some real crystal of cubic syngony for known values of the third order elastic modules were found the values of A, B, G coefficients for the reduced isotopic crystal model, matrix elements of phonons' interactions, calculated the values characterizing different phenomena in crystals for which non-linear interactions of phonons are essential with the accuracy up to numeric order of coefficients and were given tabular. As far as in this work the data on diamond crystals, which also belong to cubic syngony, we will quote the values of thermodynamic modules of elasticity for it (in terms $10^{12} dyn/sm^2$) ([9, 10]) in following table:

| $C_{11}$ | $C_{12}$ | $C_{44}$ | $C_{111}$ | $C_{112}$ | $C_{123}$ | $C_{144}$ | $C_{155}$ | $C_{456}$ |
|---|---|---|---|---|---|---|---|---|
| 10.76 | 1.25 | 5.758 | -62.6 | -22.6 | 1.12 | -6.74 | -28.6 | -8.23 |

and also the values for density $\rho = 3{,}512 \cdot 10^3$ kg/m , lattice parameter $a = 3{,}56 \cdot 10^{-10}$ m, Debaye temperature $\Theta_D^{experimental} = 2231$ K, $\Theta_D^{teoretical} = 2239$ K.

For diamond in the RICM were calculated next quantities:

$\lambda_1 = 8.488 \cdot 10^{10} N/m^2$, $\lambda_2 = 5.3568 \cdot 10^{11} N/m^2$

$V_l = 1.814 \cdot 10^4 m/s$, $V_t = 1.235 \cdot 10^4$ m/s, $\beta = 0{,}68$, $V_S = 1.346 \cdot 10^4 m/s$.

$A = -1.079 \cdot 10^{11}$ N/m$^2$, $B = -7.003 \cdot 10^{11}$ N/m$^2$, $G = -7.523 \cdot 10^{11}$ N/m$^2$,

$\bar{a} = -0.093$, $\bar{b} = -0.606$, $\bar{g} = -0.651$

## 4. The definition of SSW range of existence in reduced isotropic crystal model.

The solution of Eq.(8) to all variables we'll search in a functional form exp[i$(\Omega t - kr)$] with frequency $\Omega$ and wave vector $\mathbf{k}$. From the compatibility condition of Eq.(8) we obtain the dispersion equation for SSW [1]:

$$\Omega(\Omega^2 - W_{II}^2 \tilde{k}^2) - 2i W_{II}^2 \tilde{k}^2 \Gamma_{II} = 0 \tag{18}$$

From the relation (18) follows that $W_{II}$ is phase velocity of SSW and $\Gamma_{11}$ is attenuation coefficient of SSW. Expressions for $W_{II}$ and $\Gamma_{II}$ have simple form in the reduced isotropic crystal model [1]:

$$W_{II} = (TS^2/C\tilde{\rho})^{1/2} \quad \Gamma_{II} = \frac{r}{2\tilde{\rho}} + \left[\frac{1}{2\tilde{\rho}}\left(\frac{4}{3}\tilde{\eta} + \tilde{\zeta}\right) + \frac{1}{2C}\tilde{\kappa}\right] \tilde{k}^2 \tag{19}$$

For a phonon gas in low temperature range $\Theta_D/T \gg 1$ in the reduced isotropic crystal model are derived the next expressions:

$C = 3S = \dfrac{3V_t^2}{(2+\beta^5)T} \tilde{\rho} = \dfrac{2\pi^2}{15} \dfrac{k_B^4 T^4}{V_t^3 \hbar^3}(2+\beta^3)$, $\Theta_D = 2\pi\hbar V_S/k_B a$. SSW velocity has the form:

$$W_{II}^2 = \frac{V_t^2}{3}\left(\frac{2+\beta^3}{2+\beta^5}\right) \tag{20}$$

The values of coefficients $r$, $\tilde{k}, \tilde{\eta}, \tilde{\zeta}$ in this model are calculated in work [1].

If we work in the diffusion times by next relations [1]:

$$\tau_{\tilde{\eta}} = \frac{\tilde{\eta}}{\tilde{\rho} W_{II}^2}, \quad \tau_{\tilde{\zeta}} = \frac{\tilde{\zeta}}{\tilde{\rho} W_{II}^2}, \quad \tau_{\tilde{\kappa}} = \frac{\tilde{\kappa}}{C W_{II}^2}, \quad \tau_R = \frac{\tilde{\rho}}{r} \tag{21}$$

then value $r_{II}$ can be written as:

$$\Gamma_{II} = \frac{1}{2}\left[\frac{1}{\tau_R} + \left(\frac{4}{3}\tau_{\tilde{\eta}} + \tau_{\tilde{\zeta}}\right)\Omega^2 + \tau_{\tilde{\kappa}}\Omega^2\right] \tag{22}$$

where $\tau_N = \left\{\dfrac{4}{3}\tau_{\tilde{\eta}} + \tau_{\tilde{\zeta}} + \tau_{\tilde{k}}\right\}$.

The condition of weakly damping SSW existence

$$\Gamma_{II} \ll \Omega_{II} \tag{23}$$

leads to a known condition of SSW existence, so-called "window" [1, 3]: $\min\{\nu_{\tilde{\eta}}, \nu_{\tilde{\zeta}}, \nu_{\tilde{k}}\} \gg \Omega \gg \nu_R$, where the collision frequencies $\nu_i$ connected with diffusion times $\tau_i$ by a relationship $\nu_i = 1/\tau_i$.

Calculations of phonon gas-dynamics kinetic coefficients in the reduced isotropic crystal model relatively to the elastic modulus in [1] show that the principal role plays phonon viscosity. The number similar to the Prandtl's number in the gas theory $\Pr = C\tilde{\eta}/\tilde{\rho}\tilde{k}$ for the phonon gas for most crystals is equal to $10^2$ with a high enough accuracy. Thus one can represent Eq.(22) in the form:

$$\Gamma_{II} = \frac{1}{2}\left\{\nu_R + \frac{\Omega^2}{\nu_N}\right\}, \nu_N \equiv \frac{3}{4}\nu_{\tilde{\eta}} \qquad (24)$$

Next we will present formulas for collisions frequencies in the reduced isotropic crystal model [1] in consideration of next umklapp processes allowed by the conservation law of energy and quasi-momentum conservation law $t + t \leftrightarrow l + b_g$ and $l + t \leftrightarrow l + b_g$.

$$\nu_{\tilde{\eta}} = \frac{2\pi^2}{45}(2+\beta^3)\frac{k_B^5}{\hbar^4}\frac{T^5}{\alpha_{\tilde{\eta}}\rho V_l^2 V_t^3}$$

$$\nu_U = A_1 T^{-3} e^{-\frac{\Theta_1}{T}} + A_2 T^{-1} e^{-\frac{\Theta_2}{T}} \qquad (25)$$

$$\nu_d = C_d \frac{k_B^4 a^3}{4\pi\hbar^4 V_S^3} T^4 \left(\frac{\Delta M}{M}\right)^2 \frac{8!\zeta(8)}{4!\zeta(4)} \qquad (26)$$

where

$$A_1 = \frac{3\pi^3\beta^2(15-10\beta^2+3\beta^4)}{4(1+\beta)^6(2+\beta^5)}(1+\bar{b}+\bar{c})^2 \frac{\hbar^4 V_l^3}{k_B^3\rho a^8}, \quad \Theta_1 = \frac{2\pi\beta V_l\hbar}{k_B(1+\beta)a};$$

$$A_2 = \frac{45\pi(1-\beta)(7-5\beta)}{64(1+\beta)(2+\beta^5)\beta^2}(1+\beta^2+2\bar{b}+4\bar{c})^2 \frac{\hbar^2 V_l}{k_B\rho a^6}, \quad \Theta_2 = \frac{\pi(1+\beta)V_l\hbar}{2k_B a};$$

where $\zeta(n)$ – Riemann zeta function, $C_d = N_i/N$ – isotope concentration, $\Delta M$ – difference between atomic weight of isotopes and weight of main substance, $M$ - weight of main atoms, $D$ – sample dimension, $k_B$ –Boltzmann constant, $a_{\tilde{\eta}} \sim 10^{-2}$. The account of phonon scattering on boundaries is performed subject to the relation between mean free path at the expense of normal processes $l_N = \tau_N V_S$ and dimension of the sample D. If the relation $l_N \gg D$ is hold, then the effective mean free path of phonons coincides with the dimensions of the sample and the frequency of boundary scattering is defined as: $\nu_B = V_S/D$. SSW exist in dielectrics if realizes gas-dynamic regime in phonon description, under this condition is hold an inequality: $l_N \ll D$. Phonons in the sample depth before reaching the boundary will experience a set of normal collisions. As the result the path, which is being covered between two collisions, considerably increases. Using the well-known formulas of Brownian movement it is easy to see that trajectory between two collisions with boundary will be of the order of $D^2/l_N$ [17]. This implies that effective frequency of phonon collision with the boundary will be:

$$v_{Beff} = \frac{V_S}{D^2 v_N} = \frac{v_B}{v_N} \frac{V_S}{D} \qquad (27)$$

Further we will specify Eq.(23):

$$\Omega_{II} \geq 10 \Gamma_{II} \qquad (28)$$

from which one can get lower and upper boundary of SSW frequencies:

$$0.1 v_N - \sqrt{10^{-2} v_N^2 - v_N v_R} \leq \Omega_{II} \leq 0.1 v_N + \sqrt{10^{-2} v_N^2 - v_N v_R} \qquad (29)$$

It is apparently that

$$v_N \geq 100 v_R \qquad (30)$$

gives us in fact the window of SSW existence.

If we substitute expressions for frequencies received in RICM in Eq.(30), after elementary transformation we get the condition imposed on temperature, dimension and concentrations of isotopes

$$C_d \leq \left[ 10^{-2} v_N - v_U - \frac{V_S^2}{D^2 v_N} \right] \frac{4!\zeta(4)}{8!\zeta(8)} \frac{4\pi V_S^3}{a^3} \left( \frac{\hbar}{k_B} \right)^4 \left( \frac{M}{\Delta M} \right)^2 \qquad (31)$$

that allows us to plot the dependence of concentration $C_d$ on dimension and temperature using which one can define the temperature window of SSW existence.

## 5. SSW existence in diamond.

For execution of numeric estimations and calculations of weakly damping SSW existence condition in the diamond the following data, calculated in the reduced isotropic crystal model have been used:

$W_{II} = 0,74 \cdot 10^4$ m/s, $\Theta_1 = 984\ K$, $\Theta_2 = 1022\ K$,

$A_1 = 2,345 \cdot 10^{17}\ K^3/s$, $A_2 = 8,6 \cdot 10^{12}\ K/s$.

The expressions for collision frequencies in the reduced isotropic crystal model have the form of:

$$\begin{aligned}
v_N &= 0,2\ T^5\ s^{-1} \\
v_U &= 2{,}345 \cdot 10^{17} \cdot T^{-3} \cdot \exp[-984/T]\ c^{-1} + \\
&\quad + 8{,}6 \cdot 10^{12} \cdot T^{-1} \cdot \exp[-1022/T] \\
v_d &= 4,8 \cdot 10^3\ T^4\ C_d\ s^{-1} \\
v_{Beff} &= 9,06 \cdot 10^{12}\ D^{-2} T^5\ s^{-1}
\end{aligned} \qquad (32)$$

Substituting these expressions in Eq.(31) we get:

$C_d \leq 2.1 \cdot 10^{-7} T - 1.13 \cdot 10^9 D^{-2} T^9 - 3.29 \cdot 10^{13} \exp[-984/T]\ T^7 - 1.21 \cdot 10^9 \exp[-1022/T]\ T^5$

(33)

As far as SSW exist in the neighborhood of thermal conductivity coefficient maximum, to find out the condition of SSW existence we can also use the collision frequencies received in Callaway model while describing the experimental data on diamond thermal conductivity.

In the low temperature range in Callaway model for the description of experimental data on diamond thermal conductivity in work [8] are used next phonon collision frequencies caused by normal and umklapp processes and also processes of scattering on isotopes and boundaries of the sample:

$$v_N(k) = \frac{A_0 V_S}{2\pi} T^3 k$$

$$v_U(k) = \frac{B_0 V_S}{(2\pi)^2} T k^2 e^{-\frac{C_0}{T}} \quad (34)$$

$$v_d(k) = C_d \frac{a^3 V_S}{4\pi} \left(\frac{\Delta M}{M}\right)^2 k^4$$

where $A_0 = 7, 2 \cdot 10^{-11} K^{-3}$, $B_0 = 1, 5 \cdot 10^{14}$ m/K, $C_0 = 670K$ – phenomenological constants, calculated from the experiment [8].

Carrying out averaging procedure using the rule

$$\langle f(x) \rangle = \int_0^\infty \frac{x^4 \exp[x]}{(\exp[x]-1)^2} f(x) dx \Big/ \int_0^\infty \frac{x4 \exp[x]}{(\exp[x]-1)^2} dx \quad (35)$$

we will get next expressions for averaged frequencies of phonon collisions:

$$v_N = \frac{k_B A_0}{2\pi \hbar} T^4 \frac{5!\zeta(5)}{4!\zeta(4)}$$

$$v_U = \frac{k_B^2 B_0}{(2\pi \hbar)^2 V_S^2} T^3 e^{-\frac{C_0}{T}} \frac{6!\zeta(6)}{4!\zeta(4)} \quad (36)$$

$$v_d = C_d \frac{k_B^4 a^3}{4\pi \hbar^4 V_S^3} T^4 \left(\frac{\Delta M}{M}\right)^2 \frac{8!\zeta(8)}{4!\zeta(4)}$$

$$v_{Beff} = \frac{V_S}{D^2 v_N}$$

Similarly to previous case we substitute the expressions for frequencies received in Callaway model in Eq.(30) and get the condition imposed on temperature, dimension and concentrations of isotopes:

$$C_d \leq \left[10^{-2} v_N - v_U - \frac{V_S^2}{D^2 v_N}\right] \frac{4!\zeta(4)}{8!\zeta(8)} \frac{4\pi V_S^3}{a^3} \left(\frac{\hbar}{k_B}\right)^4 \left(\frac{M}{\Delta M}\right)^2 \quad (37)$$

The expressions for frequencies in the Callaway model have the following view:

$$v_N = 6,88 \, T^4 \, s^{-1}$$
$$v_U = 1, 4 \cdot 10^4 \, T^3 \exp[-670/T] s^{-1} \quad (38)$$

$$v_d = 4,8 \cdot 10^3 \, T^4 \, C_d \, s^{-1}$$

$$v_{Beff} = 2,63 \cdot 10^{11} \, D^{-2} T^4 \, s^{-1}$$

And the condition for concentrations of isotopes, dimension and temperature is:

$$C_d \leq 7.2 \cdot 10^{-6} - 2.3 \cdot 10^7 D^{-2} T^8 - 1.96 \exp[-670/T] T^{-1} \qquad (39)$$

In this paragraph our purpose is to exhibit the results of application of the criteria for observation of SSW (Eq.30). At the beginning we will consider the ideal case when only normal and umklapp processes are accounted. Also we consider a crystal to be unbounded and isotope pure. On Fig. 1 are presented plots of collision frequencies caused by normal and umklapp processes in the reduced isotropic crystal model and Callaway model.

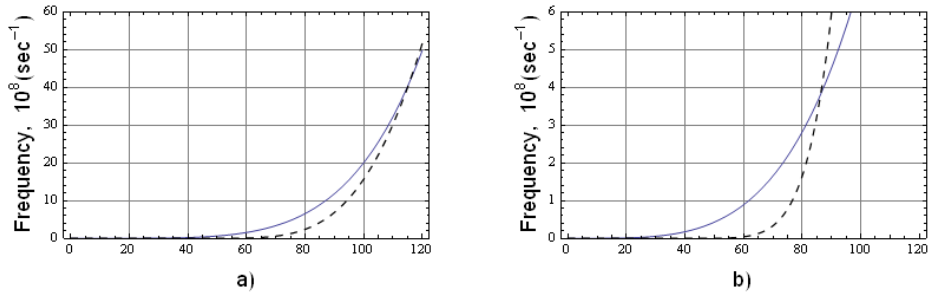

Fig. 1: N- (solid line) and U- (dash line) processes in: (a) Reduced isotropic Crystal model, (b) Callaway model

Critical temperature for SSW monitoring above which the propagation of weakly damping SSW is impossible in reduced isotropic crystal model is *T=115K* and in Callaway model - *T = 86K*, the ultimate frequencies corresponding to these temperatures are: $Q_{II} = 4.1 \cdot 10^8 s^{-1}$ and $Q_{II} = 3.8 \cdot 10^7 s^{-1}$

Using the inequality (29) we define "window" of SSW existence in RICM and Callaway model (Fig. 2). Upper curve corresponds to lower boundary and lower curve corresponds to higher boundary of SSW frequency. This figures give us an opportunity to find "temperature window" while second sound frequency is fixed and on the contrary "frequency window" while temperature is fixed.

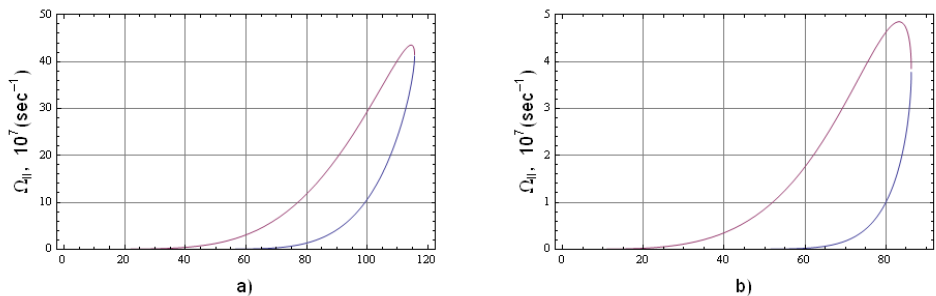

Fig. 2: Upper and lower boundaries for second sound frequency in the ideal case of N- and U-processes prevalence. (a) Reduced isotropic crystal model, (b) Callaway model

From the analysis of these plots follows that the "window" of SSW existence for different temperatures lies in different frequency intervals. In table 1 are presented "frequency windows" of SSW existence for different temperatures found in both models:

| T,K | Reduced isotropic crystal model | Callaway model |
|---|---|---|
| 40 | $3.9 \cdot 10^2 \div 4.0 \cdot 10^6$ | $2.4 \cdot 10^2 \div 3.5 \cdot 10^6$ |
| 50 | $2.8 \cdot 10^4 \div 1.2 \cdot 10^7$ | $1.3 \cdot 10^4 \div 8.6 \cdot 10^6$ |
| 60 | $4.4 \cdot 10^5 \div 3.0 \cdot 10^7$ | $2.2 \cdot 10^5 \div 1.7 \cdot 10^7$ |
| 70 | $3.1 \cdot 10^6 \div 6.4 \cdot 10^7$ | $1.7 \cdot 10^6 \div 3.1 \cdot 10^7$ |
| 80 | $1.3 \cdot 10^7 \div 1.2 \cdot 10^8$ | $1.0 \cdot 10^7 \div 4.6 \cdot 10^7$ |

In table 2 are presented "temperature windows" of SSW existence for different SSW frequencies found in both models:

Next we will take into account scattering on isotopes. Using the condition (33) we plot the dependence of critical isotope concentration on temperature under which the SSW existence is possible. The range of SSW existence will lie under received curves in both

| $\Omega_{II}, s^{-1}$ | Reduced isotropic crystal model | Callaway model |
|---|---|---|
| $10^4$ | $12 \div 47$ | $9 \div 49$ |
| $10^5$ | $19 \div 54$ | $16 \div 57$ |
| $10^6$ | $30 \div 64$ | $29 \div 67$ |
| $10^7$ | $48 \div 78$ | $52 \div 80$ |

models:

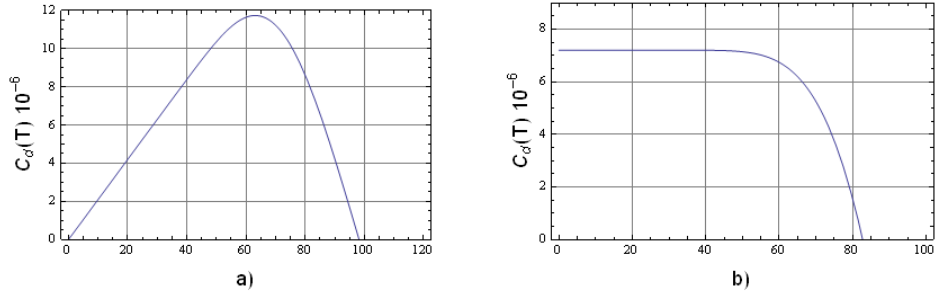

Fig. 3: The dependence of isotope concentrations on temperature: (a) Reduced isotropic crystal model, (b) Callaway model

We can see that $C_d = 2.4 \cdot 10^{-5}$ is ultimate critical concentration above which propagation of $SSW$ is impossible, corresponding critical temperature is $T=66.4K$ and ultimate frequency of $SSW$ in this point is $Q_{II} = 2.59 \cdot 10^7, s^{-1}$. Concentrations below critical $C_d$ give the "temperature window" of $SSW$ existence, which increases with concentration reduction in RICM. In Callaway model critical concentration is $C_d = 1.4 \cdot 10^{-5}$. We see that received critical concentrations are equal-order in both models.

Such difference of curves on Fig. 3 is provided by different temperature dependence of normal processes in RICM and Callaway model. As we will see further the account of boundary scattering will make these differences inessential.

Temperature window in that case for characteristic concentration of isotopes $C_d = 5 \cdot 10^{-6}$ and second sound frequency $Q_{II} = 10^7 s^{-1}$ is:

| Model | Reduced isotropic crystal | Callaway |
|---|---|---|
| Temperature window, K | 48.5 ÷ 73.8 | 54.3 ÷ 93.2 |

In next case we consider normal processes, umklapp processes and boundary scattering. From the inequality (30) follows the dependence of critical dimensions of the sample on temperature. The ranges from the right of the curves on Fig.4 are the ranges of SSW existence in RICM and Callaway model:

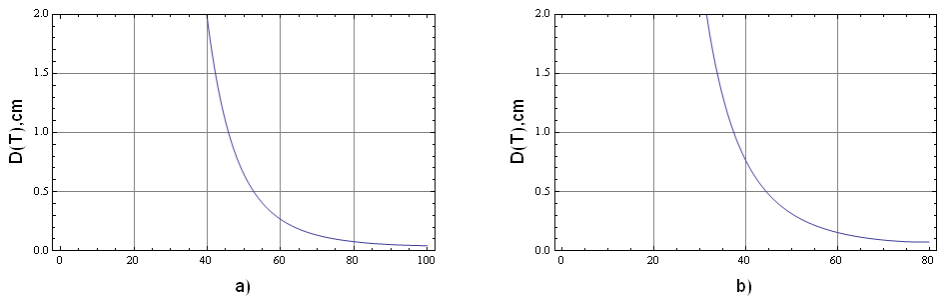

Fig. 4: The dependence of sample dimentions on temperature: (a) Reduced

isotropic crystal model, (b) Callaway model:

For example we will consider characteristic second sound frequency $Q_{II}=10^7 s^{-1}$. In order this frequency to be an eigenmode, the dimensions of crystal should be $D = 0.23 cm$, so the temperature "windows" in that case are:

| Model | Reduced isotropic crystal | Callaway |
|---|---|---|
| Temperature window, K | 47.8 ÷ 77.6 | 51.9 ÷ - |

The difference in received "windows" is provided by the same reason as in previous case and can be eliminated by the account of other resistive processes.

Further we will consider the contribution of all processes. Using the inequalities (33) and (39) we obtain the "window" of $SSW$ existence in diamond single crystal in RICM and Callaway model. On the Fig. 5, 6 are given the dependencies of concentrations $C_d$ on temperatures and dimensions. The range of $SSW$ existence lies under the plotted surfaces:

Setting definite parameters we find frequency values of weakly damping $SSW$ from the inequality (29). For example if the characteristic concentration of isotopes is $C_d = 5 \cdot 10^{-6}$, second sound frequency $Q_{II} = 10^7 s^{-1}$ and corresponding dimension of the sample is $D = 0.23 cm$, the temperature window is:

| Model | Reduced isotropic crystal | Callaway |
|---|---|---|
| Temperature window, K | 51.7 ^ 73.4 | 58.2 ^ 73.7 |

## 6. Conclusion

We believe this article demonstrates that $SSW$ can exist in diamond single crystals in the low temperature range. That was shown using two absolutely different models: the reduced isotropic crystal model, based on the elastic properties of diamond and Callaway model, based on the analysis of thermal conductivity data.

Four considered cases show how the contribution of boundaries and isotope composition of diamond influence on the range of $SSW$ existence. The account of phonon scattering on isotopes in comparison with ideal case of scattering at the expense of N- and U-processes shifts critical temperature in lower temperature range; that leads to the decrease of corresponding ultimate frequencies. Also the temperature "window" narrows and shifts in the range of higher temperatures. The account of boundary scattering considerably effects on the desired "window": narrows and shifts it in the lower temperature range. The account of all processes as against to previous cases significantly narrows the "window" and shifts it in higher temperature range.

We were able to compare our results in two models and received results appear to be satisfactory. Performed calculations of $SSW$ existence range in the reduced isotropic crystal model and Callaway model give us consistent results. Thus, it is possible to state that $SSW$ can be experimentally revealed in the diamond singe crystals under corresponding selection of parameters: isotope concentrations below critical, dimensions and temperatures of the sample.

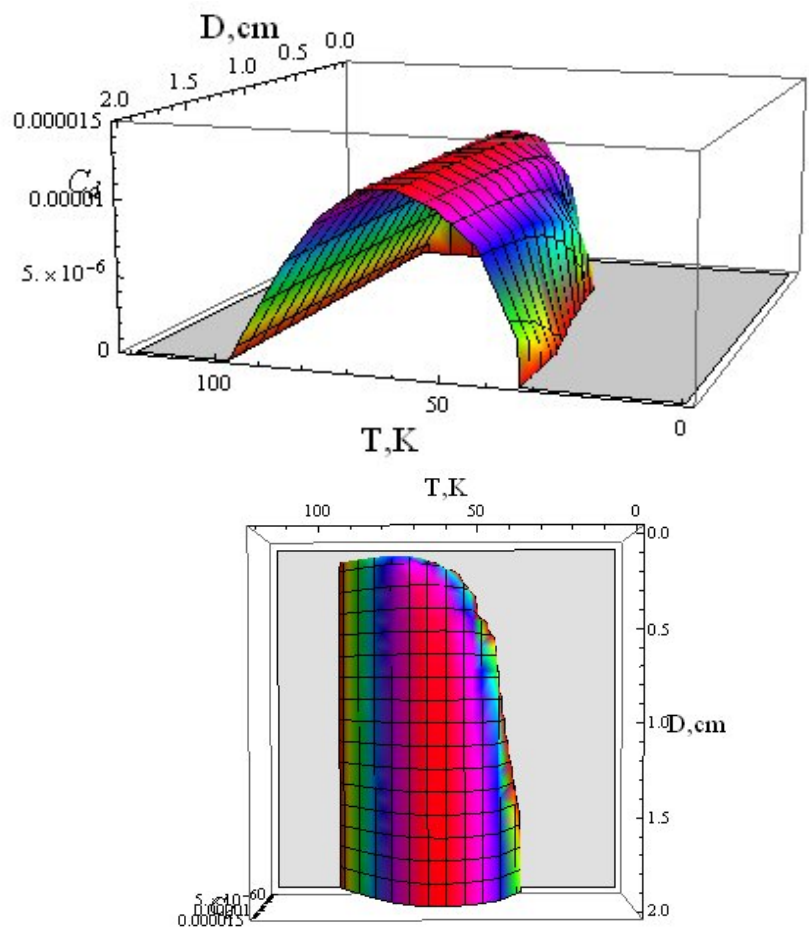

Fig. 5: The dependence of isotope concentrations on temperature and dimensions of the sample in reduced isotropic crystal model

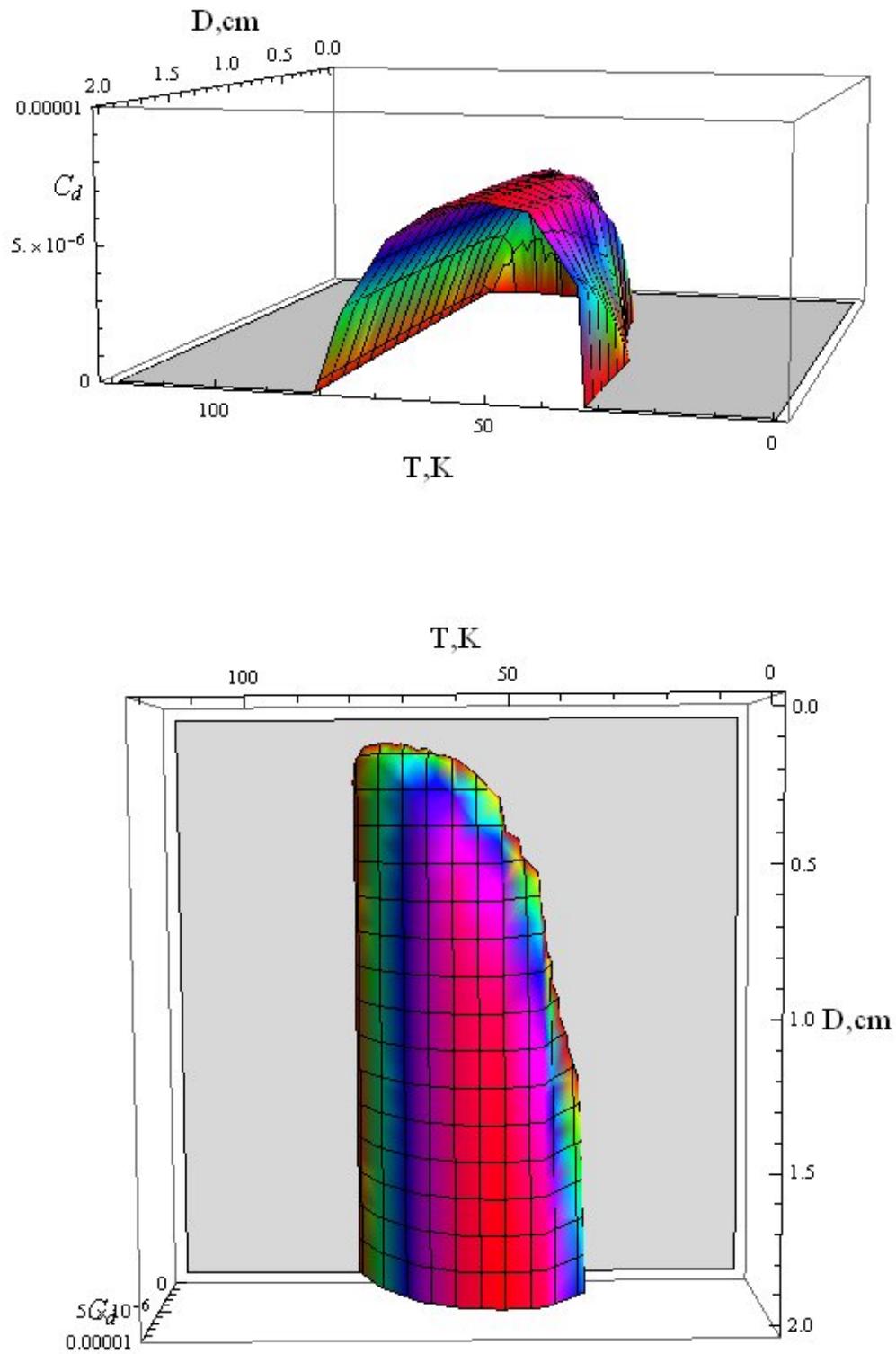

Fig. 6: The dependence of isotope concentrations on temperature and dimensions of the sample in Callaway model